\documentclass[12pt,preprint]{aastex}

\def\be{\begin{equation}}
\def\ee{\end{equation}}
\newcommand{\beqy}{\begin{eqnarray}}
\newcommand{\eeqy}{\end{eqnarray}}
\newcommand{\bce}{\begin{center}}
\newcommand{\ece}{\end{center}}

\shorttitle{skyrmion stars}
\shortauthors{Jaikumar $et~al.$}
\begin{document}

 \title{High-density skyrmion matter and neutron stars}

\author{Prashanth Jaikumar\altaffilmark{1}, Manjari Bagchi\altaffilmark{2}, and Rachid Ouyed\altaffilmark{3}}
\altaffiltext{1}{Institute of Mathematical Sciences, C.I.T Campus, Taramani, Chennai 600113 India; \\E-mail: jaikumar@imsc.res.in }

\altaffiltext{2}{Tata Institute of Fundamental Research, Homi Bhabha Road, Mumbai 400005 India}

\altaffiltext{3}{Department of Physics and Astronomy, University of Calgary, 2500 University Drive NW, Calgary, Alberta T2N 1N4 Canada}

\begin{abstract}

We examine neutron star properties based on a model of 
dense matter composed of $B$=1 skyrmions immersed in a mesonic 
mean field background. The model realizes spontaneous chiral symmetry breaking non-linearly and incorporates scale-breaking of QCD through 
a dilaton VEV that also affects the mean fields. Quartic 
self-interactions among the vector mesons are introduced on grounds of naturalness in the corresponding effective field theory. Within a 
plausible range of the quartic couplings, the model generates neutron star masses and radii that are consistent with a preponderance of observational constraints, including recent ones that point to the existence of 
relatively massive neutron stars $M\sim 1.7M_{\odot}$ and radii $R\sim$12-14 km. If the existence of neutron stars with such dimensions is confirmed, matter at supra-nuclear density is stiffer than extrapolations of most microscopic models suggest. 

\end{abstract}


\section{Introduction}
\label{sec:intro}

Neutron star astronomy, initiated by the serendipitous discovery of the first radio pulsar~\cite{Bell}, has since found over 1700 similar ``rotation-powered'' neutron stars~\cite{ATNF}.~Pulsar timing measurements in radio binaries yield a simple (unweighted) mean neutron star mass $\langle M\rangle\sim$1.4$M_{\odot}$. In contrast, measurement of general-relativistic parameters in neutron star-white dwarf binaries suggest $\langle M\rangle\sim$1.6$M_{\odot}$, owing to a few exceptionally large inferred neutron star masses in the latter case.~Examples include the binary component PSR J0621+1002~\cite{Nice} with 1.7$^{+0.10}_{-0.16}M_{\odot}$ at the 1$\sigma$ level and a more stringent value of $M\ge$1.68$M_{\odot}$ at the 2$\sigma$ level set by a neutron star binary in the Terzan 5 cluster~\cite{ransom05}.~Constraining neutron star radii is harder, due to uncertainties in atmospheric modelling and distance estimates, but bounds from thermally emitting neutron stars eg.~RXJ 1856.5-3754~\cite{WL02,Ho} imply that the canonical range of 10-12km is exceeded.~These observations have oriented attention towards 'atypical' neutron stars with large mass and possibly large radius.

An independent determination of mass (2.10$\pm0.28M_{\odot}$) and radius (13.8$\pm$1.8km) of the bursting neutron star in the low-mass X-ray binary (LMXB) EXO 0748-676 has been claimed
~\cite{ozel}, based on an accurate determination of gravitationally red-shifted  Fe and O absorption lines~\cite{Cottam}. However, the evolving nature of the source has complicated further observational tests of the same.~Furthermore, Doppler tomography of emission lines in the mass-transfer stream between the neutron star and its less massive companion suggests a more canonical value of 1.35$M_{\odot}$ for the former~\cite{Pearson}.~NASA's upcoming Constellation-X mission will improve on such spectral measurements, shrinking systematic errors on mass and radius even further.~With due caution on the observational front, the confirmation of a mass $\sim$2.0$M_{\odot}$ for a neutron star strongly constrains the equation of state of dense matter, ruling out the possibility of extreme softening at high densities.~It also implies an upper bound on the energy density of observable cold and dense matter~\cite{lp05}. A large radius $R\sim$13-14 km for a 1.4$M_{\odot}$ neutron star implies a stiff symmetry energy at densities of [1-2]$n_0$, where $n_0$ is the saturation density of nuclear matter.~These connections between the nuclear physics of dense matter and neutron star observations have been the focus of recent reviews~\cite{stein,Sed,page,lp2}.

 Uncertainties in theoretical aspects of many-body interactions at $n\hspace{-2.0mm}\gtrsim$[1-2]$n_0$ lead to predictions for the mass versus radius curve that vary widely depending on the equation of state (EoS), with the maximum mass $M_{\rm max}$ ranging from 1.4-2.7$M_{\odot}$ and radius at maximum mass $R_{\rm max}$ from 9-14km~\cite{lp}. Constraints from astrophysical observations and terrestrial laboratory data have whittled this range down~\cite{ls2006} so that most microscopic models of longstanding for nuclei and nuclear matter would struggle to explain the existence of relatively heavy neutron stars (2$M_{\odot}$) which also have a large radius ($R\hspace{-2.0mm}\sim\hspace{-2.0mm}13$km).~In fact, only few stiff equations of state eg., MPA1~\cite{mpa1}, MS0~\cite{ms} and PAL1~\cite{pal1} are consistent with the EXO 0748-676 constraint at the $1\sigma$ level, and even they fail if the matter accreted onto the neutron star is helium-rich~\footnote{The largest source of systematic error in extracting the mass and radius of EXO 0748-676 comes from the accreted mass fraction of Hydrogen $0.3<X<0.7$ (\"Ozel 2006).}.~However, while satisfying astrophysical constraints, it is also important to keep in mind constraints from laboratory data on strongly interacting matter around nuclear saturation density. This point was nicely brought out in recent papers by Li \& Steiner (2006), and
Kl\"ahn et al. (2006).

In this context, our goal in this paper is to explore further a recently proposed model of a skyrmion fluid~\cite{ob,jo}, henceforth referred to as OBJ, that was shown to lead to a very stiff equation of state and consequently generate a large maximum mass as well as radius for a neutron star. However, it was pointed out (Lattimer, private communication) that the rapid rise of the compressibility and symmetry energy just above saturation density in this model puts it at odds with experimental constraints from collective flow data~\cite{dll} and isospin diffusion studies in medium-energy heavy-ion collisions~\cite{tsang}. In this work, we determine the extent to which we can satisfy these constraints by extending the skyrmion fluid model to include higher-order interactions among the vector mesons that are theoretically motivated by arguments of naturalness in the corresponding effective field theory. The mass and radius predictions of the extended model, henceforth referred to as JOM, are also confronted with constraints set by the observation of X-ray burst oscillations, kiloHertz quasi-periodic oscillations in LMXBs and thermal emission from neutron stars. We include observational uncertainties wherever they may impact our conclusions. Our phenomenological model is able to satisfy a preponderance of these constraints.~We emphasize at the outset that our model, in its current form, has not been investigated for its applicability to nuclei or more complex phases of matter at sub-saturation densities. Our EoS presently applies only to infinite nuclear matter and neutron-rich matter in the range [1-5]$n_0$.

This paper is presented as follows. In $\S$ \ref{sec:standard}, we discuss some conventional EsoS for neutron stars; in  $\S$ \ref{sec:model} we revisit the Ouyed-Butler-Jaikumar (OBJ) model for skyrmion stars and motivate higher-order interactions that serve to tune the stiffness of the equation of state such that laboratory constraints are met. The main features of the mass versus radius curves are explained in $\S$ \ref{sec:massradius}. We compare the results obtained in the skyrmion star model to predictions of other neutron star models in light of observational bounds in $\S$ \ref{sec:observe}. Our conclusions are in $\S$ \ref{sec:conc}.

\section{Equations of State for Neutron Stars}
\label{sec:standard}

 An equation of state for dense matter is a relation between pressure and energy (or baryon) density, usually derived from an underlying microscopic model or effective theory for strong interactions. To apply to neutron stars, it should be able to generate at least a 1.4$M_{\odot}$ static neutron star with a radius in the 10-14km range. The connection to the underlying microscopic theory can be formulated in several ways: examples include relativistic mean field theory, non-relativistic potential models, relativistic Dirac-Brueckner-Hartree-Fock theory $etc$~\cite{ab,lp}. As we are concerned with recent findings of relatively heavy neutron stars, we consider three stiff eos: MS0~\cite{ms}, APR~\cite{apr} and UU~\cite{Wiringa}.

(i) M\"uller \& Serot (1996) used a relativistic theory of point-like nucleons interacting via mesonic degrees of freedom. These are the neutral scalar ($\sigma$) and vector ($\omega$) fields, plus the isovector $\rho$ meson. In this model, like or unlike meson-meson interactions are encoded by terms that are polynomials in the fields. By demanding a match to the properties of nuclear matter at saturation, they obtained a sequence of EsoS that depend on the coupling constants of the polynomial interactions. The stiffest eos (MS0) corresponds to vanishing couplings and yields a maximum mass of $2.7M_{\odot}$. This model is consistent with a large neutron star mass and radius$~\sim$14km for static configurations. Presently, the 1$\sigma$ limits on the radius of the bursting neutron star source EXO 0748-676 are 13.8$\pm$1.8km. However, there is no fundamental symmetry principle that requires the higher-order couplings to vanish. Introducing natural values for these couplings drastically reduces the maximum mass to $\leq 2.0M_{\odot}$.

(ii) Akmal $et~al.$ (1998) obtained the APR EoS based on the Argonne $\upsilon_{18}$ nucleon-nucleon interaction
~\cite{Stoks}, Urbana IX three-nucleon interaction~\cite{Pud} and a relativistic boost term
~\cite{Forest} as microscopic input. This EoS gives a maximum neutron star mass of 2.2$M_{\odot}$ and does not lie within the 1$\sigma$ limits on the mass-radius estimate of the neutron star in EXO 0748-676, even assuming the accreted matter is mostly hydrogen.

(iii) EoS UU is obtained via similar variational methods applied to an older two-nucleon and three-nucleon interaction (Urbana $\upsilon_{14}$+UVII). This model yields a maximum mass of $2.2~M_{\odot}$ at radius 10 km~\cite{Wiringa}. However, a radius larger than about 11km is not supported by this equation of state for a static neutron star with mass greater than 1.4$M_{\odot}$.

Our purpose in selecting and highlighting these EsoS is two-fold. Firstly, these models are representative of complementary philosophies behind constructing an equation of state for dense matter: as in (i) forgo the connection to laboratory data on vacuum two-nucleon interactions and focus instead on the empirical properties of large nuclei and infinite nuclear matter within a relativistic theory; or as in (ii) and (iii) insist on a satisfactory description of available data on the structure and interaction of few nucleon systems (free or bound) in a non-covariant approach. Secondly, these are among the stiffest equations of state that arise from hadronic degrees of freedom alone. It is possible that hybrid equations of state that allow for quark matter at high density can be almost just as stiff~\cite{Alf}, but we will not consider quark matter EsoS in this work.

These three EsoS are used in our neutron star mass-radius plots (Figs.\ref{fig:mrobs1},\ref{fig:mrobs2}) and compared with the EsoS for dense skyrmion matter which we now describe.

\section{The Skyrmion fluid}
\label{sec:model}

\subsection{Nuclear Matter Phenomenology}

Before the advent of Quantum Chromodynamics (QCD), T. H. R. Skyrme proposed a description of baryons as topological solitons in a mesonic field theory that realizes spontaneous chiral symmetry breaking in non-linear fashion~\cite{sky}. This model is now qualitatively supported by studies of large $N_c$ QCD which suggest that mesonic degrees of freedoms are fundamental and baryons arise as solitons. When augmented by the inclusion of low-lying vector mesons ($m_V\le$1GeV) and flavor symmetry breaking effects, the Skyrme model can provide a reasonable description of static baryon properties such as mass splittings, charge radii and magnetic moments~\cite{Schecter}.~In the 2-nucleon sector, the problem of finding a sufficiently attractive isoscalar central and spin-orbit force within the Skyrme model at distances 1fm$\le r\le$2fm has held up progress. One solution is to include a dilaton field that mocks up scale-breaking in QCD and provides attraction in these channels.

To make progress towards an equation of state for a skyrmion fluid, K\"albermann (1997) introduced a self-consistent model that incorporates medium effects through the response of the dilaton\footnote{Since chiral symmetry is broken non-linearly,the usual $\sigma$ field as the chiral partner of the pion does not appear in the theory.}~$\sigma$ and the isoscalar $\omega$-field to a smooth density distribution obtained by integrating over the collective co-ordinates of the Skyrmion. This is equivalent to an ensemble of non-interacting $B$=1 Skyrmions, a valid picture upto a separation of 0.8fm ($n\lesssim$5$n_0$). A subsequent work~\cite{jo} extended this model to asymmetric matter by incorporating the $\rho$ meson in the spirit of standard mean field approaches. The Lagrangian for the Skyrme model, augmented by the $\sigma,\omega$ fields, and including isospin-breaking effects from the $\rho$ meson as well as explicit scale-breaking effects from the dilaton and quark masses is given in Jaikumar \& Ouyed (2006). A fit to nuclear matter phenomenology is achieved through additional parameters in the dilaton potential, which is given by~\cite{kal}

\begin{eqnarray}
\label{anomalyeq}
V(\sigma)&=&B[1+{\rm e}^{4\sigma}(4\sigma-1)+a_1({\rm e}^{-\sigma}-1)\\ \nonumber
&&+a_2({\rm e}^{\sigma}-1)+a_3({\rm e}^{2\sigma}-1)+a_4({\rm e}^{3\sigma}-1)]\,,
\end{eqnarray}
where $B\approx (240~{\rm MeV})^4$ is related to the Bag constant (the non-perturbative glue that breaks scale-invariance in QCD). The six unknown parameters of the model are $a_1$-$a_4$, $g_w$ the $\omega$-$N$ coupling and $g_{\rho}$, the $\rho$-$N$ coupling.  To determine the $a_i$'s, the following constraints are imposed: the scale anomaly condition ${dV_{\sigma}\over d\sigma }|_{\sigma=0} = 0$ which implies that $a_1 = a_2 + 2a_3 + 3a_4$, the stationarity w.r.t $\sigma_0$, viz., $\partial E/\partial\sigma_0=0$, a binding energy/nucleon of -16 MeV for infinite nuclear matter at saturation density ($n_0=0.16$fm$^{-3}$), and a choice of the compressibility $K$. $E$ is the energy density of the fluid and $\sigma_0$ is the non-vanishing mean field value of the time component of the $\sigma$ field.  At saturation, $\sigma_0$ is determined by the choice of the effective mass $M$, which then also fixes $g_w$. The choice of symmetry energy and effective mass at saturation fixes $g_{\rho}$. Once the $a_i$'s are determined, $\sigma_0$ is generally obtained from its equation of motion for an arbitrary density. The $a_i$'s show very weak dependence on the choice of $K$ in the range 200-300 MeV, while displaying more sensitivity to the choice of effective mass. Without further modifications, the model displays a sharp rise in the compressibility just above saturation, rising from $K$=$240$ MeV (our choice) to $K\sim 2000$ MeV for a 10\% increase in baryon density. Similarly, the symmetry energy rises too steeply in this range to be consistent with experimental constraints (see $\S$\ref{symmeng}). These inconsistencies are a consequence of the specific form of the dilaton potential, which is essential to preserve the trace anomaly relation (scale-breaking). Therefore, a modification of $V(\sigma)$ is not desirable. It is the exponential sensitivity of the curvature of the potential $V(\sigma)$ to the dilaton VEV that drives the compressibility to large values. One way to address this issue is to view the Skyrme model in the mean field approximation as an effective field theory of hadrons, so that the Lagrangian can be extended to include higher-order terms (meson self-interactions) that parameterize unknown physics at a more microscopic level. This rationale is also employed in M\"uller \& Serot (1996) although their model has an explicit $\sigma$ meson, point-like nucleons and no scalar-vector mixing, while our model has a dilaton as the only scalar, and includes scalar-vector mixing.~If the meson fields are viewed as relativistic functionals, the higher-order interactions can be thought of as parts of an effective potential that determines their mean field values at a particular density through the stationarity of the effective action associated to the Skyrme lagrangian~\cite{Furn}. They therefore modify the density dependence of the meson fields as well as the properties of the background skyrmion fluid that couples to these fields.~We restrict ourselves to quartic self-interactions in the $\rho$ and $\omega$ fields with coupling constants whose value can be surmised by naturalness. Then, higher-order terms such as six or eight-meson self-interactions do not substantially change the results obtained in the quartic case. In the next section, we implement this procedure to obtain an acceptable behaviour of the compressibility and symmetry energy in dense Skyrmion matter.

\subsection{Mean-field equations}

The additional quartic interactions take the form 

\beqy
{\cal L}_{\omega}&=&{\cal L}_{\omega}^0+{\cal L}_{\omega}^{\rm int}; \quad{\cal L}_{\omega}^{\rm int}=+\frac{\xi}{4!}g_{\omega}^4\omega^4\,,\\
{\cal L}_{\omega}^0&=&-\frac{1}{4}F_{\mu\nu}^{\omega}F^{\omega,\mu\nu}+\frac{1}{2}{\rm e}^{2\sigma}m_{\omega}^2\omega^2-g_{\omega}n\omega
\eeqy
for the $\omega$ field and
\beqy
{\cal L}_{\rho}&=&{\cal L}_{\rho}^0+{\cal L}_{\rho}^{\rm int};\quad {\cal L}_{\rho}^{\rm int}=+\frac{\chi}{4!}g_{\rho}^4\rho^4\,,\\
{\cal L}_{\rho}^0&=&-\frac{1}{4}F_{\mu\nu}^{\rho}F^{\rho,\mu\nu}+\frac{1}{2}{\rm e}^{2\sigma}m_{\rho}^2\rho^2-\frac{g_{\rho}}{2}n_I\rho
\eeqy
for the $\rho$ field. Here, the $F$'s are the standard field-strength tensors for the abelian ($\omega$) and non-abelian cases ($\rho$). Working at $T=0$, we have
\be
n=\frac{\left(k_{F_p}^3+k_{F_n}^3\right)}{3\pi^2}\,, n_I=\frac{\left(k_{F_p}^3-k_{F_n}^3\right)}{3\pi^2}\,,
\ee
where $k_{F_i}$ is the Fermi momentum of species $i=n{\rm eutron},
p{\rm roton}$. Then, the Skyrme Lagrangian is compactly expressed as
\beqy
{\cal L}&=&{\cal L}_2 + {\cal L}_4 + {\cal L}_{\omega}+{\cal L}_{\rho} - V(\sigma)\,,
\eeqy
where ${\cal L}_2,{\cal L}_4$ involve gradients of the Skyrmion profile. For the densities of interest ($n_0\leq n\leq 5n_0$) where the approximation of non-overlapping skyrmions is valid, this profile drops off fast enough that the mean-field averaging simply counts the number of individual Skyrmions in a given volume, equivalent to a non-interacting Fermi gas model. At $n\geq 5n_0$, we expect corrections to our mean field model from Skyrmion overlap. We also do not expect such a mean field treatment to apply at densities much {\it lower} than saturation density, since Skyrmions do not form a uniform fluid there. Thus, our model is restricted to ($n_0\leq n\leq 5n_0$). For the $\omega$ and $\rho$ fields, the equations of motion read as follows:
\beqy
\label{cubic}
&&a_{\omega}\omega_0^3+b_{\omega}\omega_0+c_{\omega}=0;\\ &&a_{\omega}=\frac{\xi g_{\omega}^4}{6}, b_{\omega}={\rm e}^{2\sigma_0}m_{\omega}^2,\,c_{\omega}=-g_{\omega}n\,.\\
&&a_{\rho}\rho_0^3+b_{\rho}\rho_0+c_{\rho}=0;\\ 
&&a_{\rho}=\frac{\chi g_{\rho}^4}{6},\,b_{\rho}={\rm e}^{2\sigma_0}m_{\rho}^2,\,c_{\rho}=\frac{g_{\rho}n\delta}{2}\,,
\eeqy
where $\omega_0,\rho_0$ denote mean field values and $\delta$=(1-2$x$), with $x$=($n_I$-$n$)/2$n$ being the proton fraction of neutron-rich matter ($x$=$1/2$ for symmetric matter). The magnitudes~\footnote{We choose the sign of the couplings to be positive since this guarantees zero mean fields at vanishing source density (baryon/isospin).} of the quartic couplings $\xi,\chi$ are estimated from the naturalness argument for an effective field theory, viz., that the co-efficients of the various terms in the Lagrangian, through a given order of truncation, should be of the same size when expressed in an appropriate dimensionless form. Thus, we find
\be
\xi g_{\omega}^2\sim 12\left(\frac{{\rm e}^{\sigma_0}m_{\omega}}{M_0}\right)^2;\quad \chi g_{\rho}^2\sim 192\left(\frac{{\rm e}^{\sigma_0}m_{\rho}}{M_0}\right)^2 \,.
\ee 

We choose the effective mass, given by $M={\rm e}^{\sigma_0}M_0$, to be 600 MeV at saturation density, so that ${\rm e}^{\sigma_0}$=2/3 for a bare nucleon mass $M_0$=900 MeV (neglecting the $\sim$40 MeV contribution from explicit symmetry breaking). Since the model is fit to saturation properties, $g_{\omega}$ itself depends on $\xi$, whose value must be chosen so as to satisfy the naturalness condition above. Furthermore, real solutions to the equation of motion for the dilaton Eqn.(\ref{vsigma}) cease to exist beyond a small range of couplings. This restricts us to $0.1\leq\xi\leq 0.3$ and $1.0\leq\chi\leq 2.0$. These values differ from those in M\"uller \& Serot (1996) due to additional factors of ${\rm e}^{2\sigma_0}$ from the dilaton (metric) that appear in the fitting expressions for $g_{\omega}$ and $g_{\rho}$ and the qualitatively different form of the dilaton potential (it contains all powers in $\sigma$).~The energy densities corresponding to the vector mean fields are
\be
E_{\omega}=-\frac{a_{\omega}\omega_0^4}{4}-\frac{b_{\omega}\omega_0^2}{2}-c_{\omega}\omega_0\,,\,E_{\rho}=-\frac{a_{\rho}\rho_0^4}{4}-\frac{b_{\rho}\rho_0^2}{2}-c_{\rho}\rho_0\,,
\ee
so the total energy density is then
\beqy
\label{energy}
&&E=E_{\rm kin}+E_{\omega}+E_{\rho}+V(\sigma)\,,\\
&&E_{\rm kin}=\sum_{n,p}\left[\frac{k_FE_F(E_F^2+k_F^2)}{8\pi^2}-\frac{M^4}{8\pi^2}{\rm ln}\left(\frac{k_F+E_F}{M}\right)\right]\,,\nonumber
\eeqy
where the kinetic energy $E_{\rm kin}(k_F)$ comes from a 
Lorentz boost of the static Skyrmion to momentum $k_F$~\cite{kal}.
In Eqn.(\ref{energy}), $E_{F_i}=\sqrt{k_{F_i}^2+M^2}$ since the Dirac effective mass is the same for both neutrons and protons. The binding energy is $E/n-M_0$ and the pressure is given by
\be
\label{pressure}
P=n\left[0.5\left(\sum_{n,p}E_F\right)+g_{\omega}\omega_0-\frac{g_{\rho}\rho_0\delta}{2}\right]-E\,.
\ee

The effective mass $M$ is determined at any density from the equation of motion for $\sigma_0$

\beqy
\label{vsigma}
\left[\sum_{n,p}\frac{k_F^3E_F}{4\pi^2}\right]+4E_{\rm kin}+\frac{dV}{d\sigma_0}-b_{\omega}\omega_0^2-b_{\rho}\rho_0^2=0\,.
\eeqy 

\subsection{Compressibility}

For symmetric matter, $\delta=0$ and $\rho_0$ vanishes. The compressibility is defined as

\be
K=9\frac{dP}{dn} = 9 \left[n^2 \frac{\partial ^2 (E/ n)}{\partial n^2}+2n \frac{\partial (E/ n)}{\partial n} \right]\,.
\ee 

In the present case, this is equivalent to

\begin{eqnarray}
&&K=9n\left[\frac{\partial ^2E}{\partial n^2}-\left(\frac{\partial^2 E}{\partial n\partial\sigma}\right)\frac{d\sigma}{dn}\right];\\
&&\frac{d\sigma}{dn}=\left(\frac{\partial^2 E}{\partial n\partial\sigma}\right)/\left(\frac{\partial ^2E}{\partial\sigma^2}\right)\,,\nonumber\\ \label{Enn}
&&\frac{\partial ^2E}{\partial n^2}=\frac{k_F^2}{3nE_F}+\frac{g_{\omega}^2}{b_{\omega}+3a_{\omega}\omega_0^2}\,,\\ 
&&\frac{\partial^2 E}{\partial n\partial\sigma}=\frac{M^2}{E_F}-\frac{2g_{\omega}b_{\omega}\omega_0}{b_{\omega}+3a_{\omega}\omega_0^2}\,,\\ \label{Ensig}
&&\frac{\partial^2E}{\partial\sigma^2}=\left[\frac{3nM^2}{E_F}-4\frac{dV}{d\sigma_0}+\frac{d^2V}{d\sigma_0^2}+6b_{\omega}\omega_0^2
\frac{(b_{\omega}+a_{\omega}\omega_0^2)}{(b_{\omega}+3a_{\omega}\omega_0^2)}\right]\,.\nonumber
\end{eqnarray}

For the choice $K$=240 MeV, binding energy=-16 MeV and $M$=600 MeV at saturation density, the best fit $a_i$ values for various $\xi$ are listed in Table~1, correcting an unfortunate error in their values quoted for $\xi$=0 in Jaikumar \& Ouyed (2006).

\begin{figure}
\epsscale{0.7}
\plotone{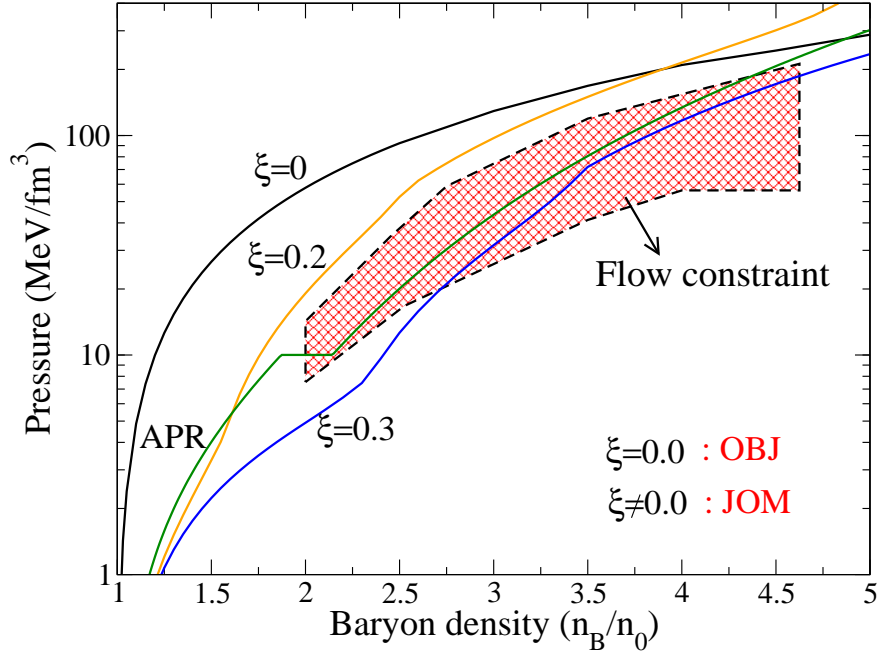}
\vskip 2.0cm
\plotone{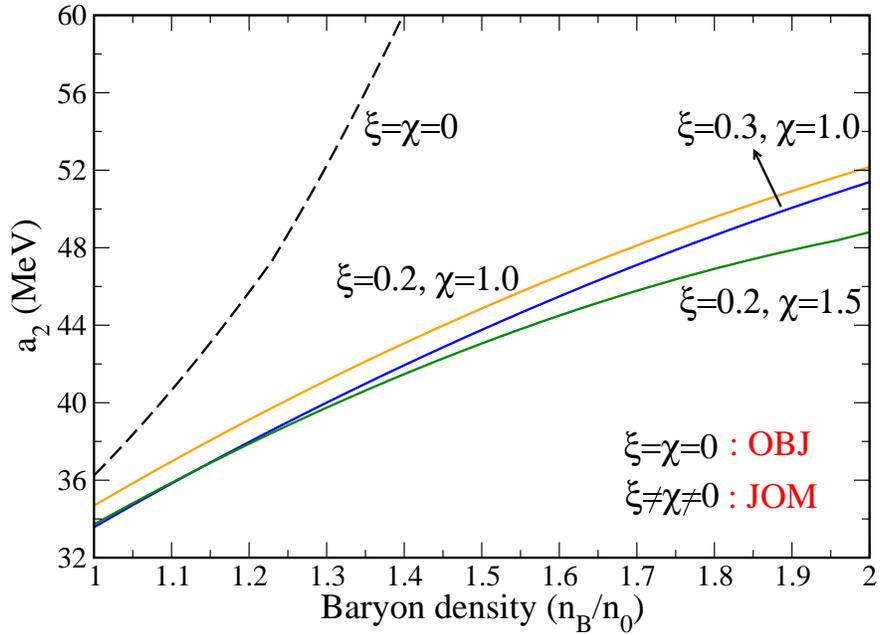}
\caption{Upper panel: EoS of symmetric matter for various values of the quartic coupling $\xi$ in the skyrmion fluid model. The APR EoS is also shown. The hatched region is the flow constraint~\cite{dll}. Lower panel: Symmetry energy $a_2$ for natural values of the couplings $\xi$ and $\chi$ in the same model.\label{fig:compress}}
\end{figure}

For $\xi$=0, at saturation, freedom in choosing the fit parameters $a_i$ allows for a delicate cancellation between various contributions to the compressibility such that $K$ is small ($\sim$200-300 MeV). This is achieved largely as a result of fine-tuning $d^2V/d\sigma^2$ away from its natural scale$~\sim B$. As we move to slightly higher density, this fine tuning cannot be recovered since $V(\sigma_0)$ is exponentially sensitive to changes in the $\sigma$ VEV. Consequently, the contribution from the $\omega$-meson appearing in the $\partial ^2E/\partial n^2$ term dominates, pushing the compressibility to unnaturally large values, as in the OBJ model. For $\xi\neq 0$, $a_{\omega}\neq0$ and $\omega_0$ being large, it is clear from Eqn.(\ref{Enn}) that $\partial ^2E/\partial n^2$ and hence $K$ is substantially reduced, as in the JOM model. This effect is reflected in the EoS of symmetric matter, shown in the upper panel of Fig.\ref{fig:compress}. By definition, the slope of the pressure-density curve is proportional to the compressibility. The hatched region represents constraints on the EoS of symmetric matter from the analysis of collective flow data in nucleus-nucleus collisions at (1-2)GeV/nucleon~\cite{dll}. While the $\xi=0$ curve (OBJ) is clearly too stiff, choosing natural values of the coupling ($0.1\leq\xi\leq 0.3$) provides a considerable improvement, and the JOM model is able to satisfy the flow constraint for $\xi\sim 0.25$. The APR EoS, which has a
phase transition at $n\sim$2$n_0$, is also shown for comparison.

\clearpage
\clearpage

\subsection{Symmetry energy}
\label{symmeng}
In neutron-rich matter, $\delta\neq 0$ and the energy/particle $\epsilon=E/n$ can be expanded about the symmetric point $\delta=0$ ($x=1/2$) :
\be
\epsilon(n,\delta)=\epsilon(n,0)+a_2(n,\sigma_0(n))\delta^2+a_4(n,\sigma_0(n))\delta^4+...\,\label{esym}
\ee
where ... are higher order terms, and
\beqy
&&a_2(n,\sigma_0(n))=\frac{1}{2!}\left(\frac{\partial^2\epsilon}{\partial \delta^2}\right)_{\delta=0}=\frac{g_{\rho}^2p_F^3}{12\pi^2m_{\rho}^2{\rm e}^{2\sigma_0}}+\frac{1}{6}\frac{p_F^2}{E_F}\,,\nonumber\\
&&a_4(n,\sigma_0(n))=\frac{1}{4!}\left(\frac{\partial^4\epsilon}{\partial \delta^4}\right)_{\delta=0}=\nonumber\\
&&\frac{k_F^6}{648E_F}\left[\frac{4}{k_F^4}+\frac{3}{k_F^2E_F^2}+\frac{3}{E_F^4}\right]-\left(\frac{\chi}{384}\right)\frac{n^3}{(b_{\rho}/g_{\rho}^2)^4}\,.
\eeqy

Studies of neutron matter show that the value of $x$ for beta-equilibrated matter obtained by retaining only $a_2$ is a good approximation over a wide range of $n$~\cite{pal2}. Recently, Steiner (2006) has examined the role of the quartic coefficient $a_4$ at high density, finding the effects on the EoS to be small, although the threshold for the onset of rapid neutrino cooling via the direct urca process can change considerably. As we are focusing on the equation of state in this work, we drop the $a_4$ term, keeping in mind that were we to retain the $a_4$ term in Eqn.(\ref{esym}), the correction to the energy difference of symmetric and neutron-rich matter is at the 5-10\% level for densities of interest. Proceeding with $a_2$ alone, as defined in  Eqn.(\ref{esym}), beta equilibrium and charge neutrality conditions yield
\be
\label{betaq}
(3\pi^2nx)^{1/3}=4a_2\delta\quad ,
\ee
which fixes a solution $x=x_0(\sigma_0)$ for any baryon density. Inverting this relation, and solving Eqn.(\ref{betaq}) along with the mean field equations for $\sigma_0,\omega_0$ and $\rho_0$, we determine $x_0(n)$ explicitly. For the density range $n/n_0=[1..5]$, $0.045\leq x_0\leq 0.11$ for $\xi,\chi\neq 0$ and the direct urca threshold is just about reached at the upper limit. This conclusion could be affected by the inclusion of the $a_4$ term in the expansion of Eqn.~(\ref{esym}).

The lower panel of Fig.\ref{fig:compress} shows the density-dependence of the symmetry energy $a_2(\sigma_0(n),n)$. Since $a_2$ in Eqn.(\ref{betaq}) depends on $\sigma_0$, which takes different values for the same density in symmetric and asymmetric matter, the curves do not begin at 32 MeV (our choice for symmetric matter). We have chosen $a_2$ from Eqn.(\ref{betaq}) to represent the symmetry energy since it is this quantity that determines the proton fraction of beta-equilibrated matter. This is different from the usual identification of the symmetry energy, which is made at $\delta=0$. In addition, it must be noted that $\sigma_0$ and hence $a_2$ now depend on $\chi$ as well as $\xi$. For $\xi$=$\chi$=0, a stiff symmetry energy results, that behaves approximately as $a_2(n)$=$a_2(n_0)\left(n/n_0\right)^{1.26}$in the range [1-1.5]$n_0$. Li \& Steiner (2006) have argued that such a parametrically stiff symmetry energy is inconsistent with isospin-diffusion data in heavy-ion collisions~\cite{Rami} and the measured neutron-skin thickness of Lead~\cite{star}.~For natural values of the couplings, the symmetry energy softens considerably, behaving approximately as $a_2(n)$=$a_2(n_0)\left(n/n_0\right)^{0.71}$, which 
is consistent with the aforementioned experimental studies. While increasing
$\xi$ does not change the density dependence, increasing $\chi$ softens the symmetry energy at high density. This is a consequence of scalar-vector mixing in our model.

\begin{figure}[!t]
\epsscale{0.8}
\vskip 0.75cm
\plotone{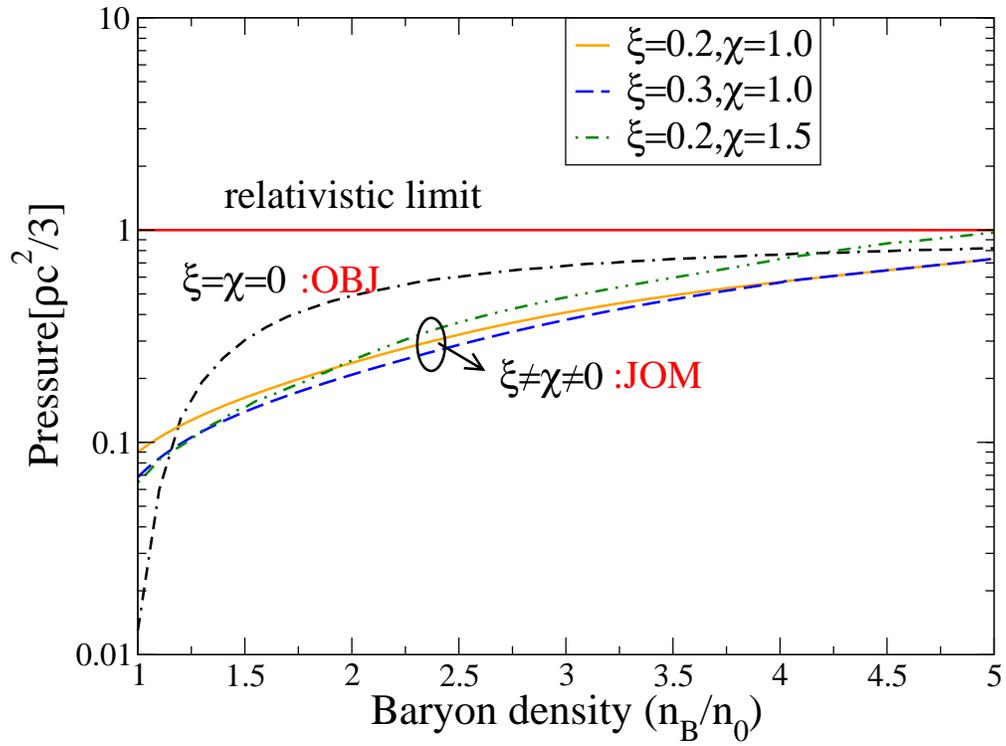}
\caption{Pressure of beta-equilibrated matter (relative to a relativistic gas of the same energy density) as a function of density for the OBJ and JOM EsoS.\label{fig:pressure}}
\end{figure}

The pressure of neutron-rich matter in $\beta$-equilibrium (excluding lepton pressure), scaled to that of a relativistic fluid with the same energy density, is shown in Fig.~\ref{fig:pressure}. Close to saturation, the pressure is well-described by $P(n,\delta)=P_{\rm sym}(n)+n^2\delta^2da_2/dn$ where $P_{\rm sym}$ is the pressure of symmetric matter. The curves in the lower panel of Fig~\ref{fig:compress} imply that $\delta$ is smaller while $da_2/dn$ is larger for vanishing couplings, as compared to the case with non-zero couplings. This drives the initial rapid rise of the pressure relative to the energy density for $\xi$=$\chi$=0. For non-zero couplings, the softer symmetry energy and larger value of $\delta$ 
lead to a larger pressure at saturation but a more gradual increase relative
to the energy density. Increasing $\xi$ at fixed $\chi$ softens the equation of state and has a progressively decreasing effect at high densities, where the kinetic contribution to the pressure begins to dominate; hence the gradual approach to the relativistic limit. Increasing $\chi$ at fixed $\xi$ has a more dramatic effect on the pressure and energy density, causing the EoS to exceed the relativistic limit at high densities. This is due to the dynamics dictated by the dilaton potential at high density and scalar-vector mixing in our model. The nucleon's effective mass increases rapidly with increasing density beyond a certain value of $\chi$ and the skyrmion shrinks, akin to a strong repulsive force.

\clearpage
\clearpage

\section{Mass-Radius relation}
\label{sec:massradius}

\begin{figure}
\epsscale{1.0}
\vskip 0.75cm
\plotone{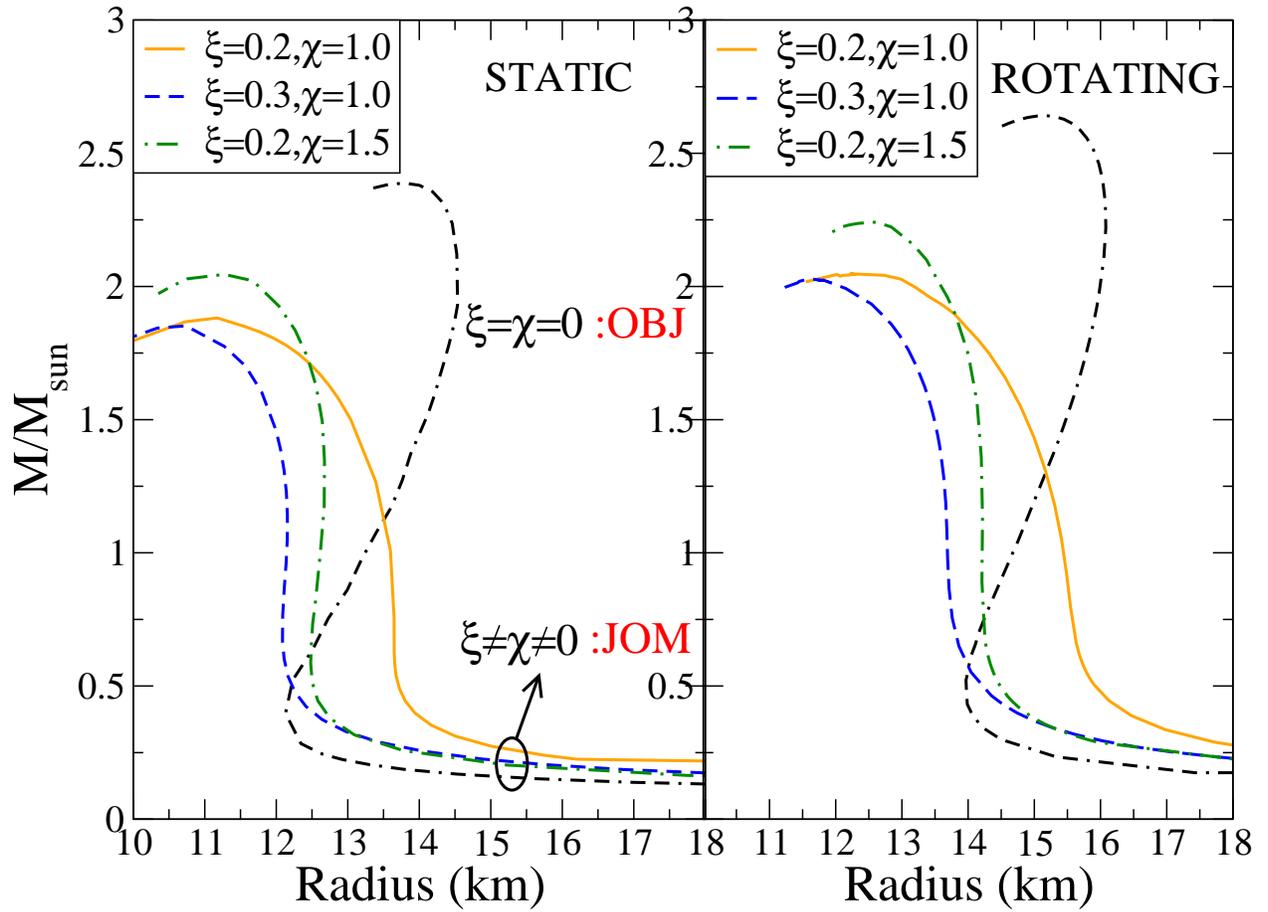}
\caption{Mass-radius curves for static (left panel) and rotating (right panel) stars with the OBJ and JOM EsoS.\label{fig:massradius}}
\end{figure}

We employ the $RNS$ code~\cite{Ster95,Ster98} to generate  mass versus radius curves for a sequence of static and rapidly rotating neutron stars with the OBJ and JOM EsoS. In both cases, the composition of the star is as follows: (i) 1$\leq $$n/n_0$$\leq$5: $(n,p,{\rm e}^-)$ matter in beta-equilibrium with nucleonic pressure and energy density given by Eqns.(\ref{energy}) and (\ref{pressure}); (ii) $n/n_0\leq$1: the BBP equation of state~\cite{BBP} for densities below nuclear saturation density, matched to the BPS equation of state~\cite{BPS} for the low-density nuclear crust of the star. For non-zero couplings (JOM EoS), the maximum mass is not reached until densities larger than $5n_0$, so we had to extend our model to higher densities. Therefore, the maximum masses for JOM EoS indicated in Fig.\ref{fig:massradius} are to be viewed as extrapolations into a regime where the approximation of non-overlapping $B$=$1$ spherically symmetric skyrmions is not likely to hold. A more accurate method would have to first determine the topology and size of deformed or overlapping skyrmions, which we do not attempt in this work.

\clearpage
\clearpage

The variation in the mass-radius curves with the values of $\xi$ and $\chi$ reflects the following two facts: the stiffer the EoS at supra-nuclear densities, the larger the maximum mass and (ii) the larger the pressure in the range [1-2]$n_0$, the larger the radius for a (1-2)$M_{\odot}$ star. These correlations have been established and emphasized in previous work~\cite{lp}. The extreme stiffness of the symmetry energy for $\xi$=$\chi$=0 results in large pressures in the range [1.5-2.0]$n_0$ and leads to a large radius $R\sim 14$km for a 1.4$M_{\odot}$ star in the OBJ model. The softening induced by non-zero couplings lowers the pressure in this range, leading to smaller radii in the JOM model. The maximum mass is also lower with respect to the case when the couplings are set to zero. For the rapidly rotating models, we chose a rotation frequency $\nu\sim$600Hz, corresponding to a period of 1.6ms. In general, the additional centrifugal forces in a rotating star help to counteract the pull of gravity, resulting in larger radii for a given mass.

As pointed out in the context of a different mean field model (M\"uller \& Serot 1996), the inclusion of the additional quartic terms softens the equation of state for beta-equilibrated neutron-rich matter considerably, making it difficult to obtain a neutron star mass larger than 2$M_{\odot}$. Within our model, while a large radius and mass is possible if these quartic terms are omitted, the requirement of respecting laboratory constraints near saturation density and arguments of naturalness imply that the inclusion of these couplings is essential and its consequences (a lowering of maximum mass and radius) quite general. We now turn to compare our results (along with those from other EsoS) with some current observational bounds on the mass and radius of neutron stars.

\clearpage
\begin{figure*}
\plotone{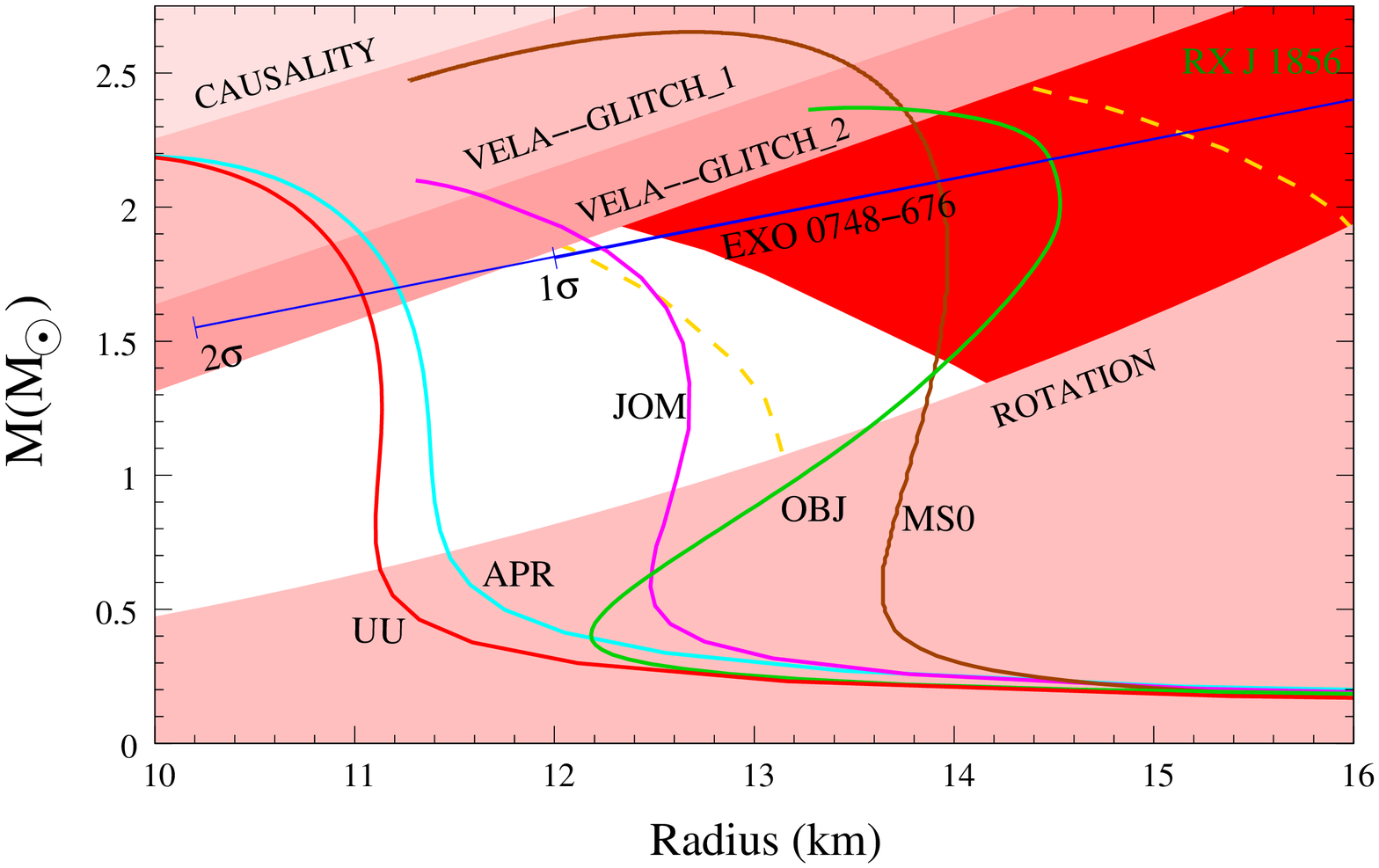}
\caption{Observational constraints on neutron stars: The region allowed by observations of RXJ 1856.5-3754~\cite{WL02} is shown in red; regions excluded by glitches in the Vela pulsar~\cite{Link}, by the spin-rate of J1748-2446ad~\cite{Hess}, and by causality are demarcated in pink. Mass-radius curves for static stars corresponding to some stiff equations of state (APR, UU, MS0) are plotted along with the ones in this work (OBJ, JOM). The z=0.35 constraint for EXO 0748-676 rules out the APR and UU equations of state at the 1$\sigma$ level but allows them at the 2$\sigma$ level. Yellow dashed lines are 90\% confidence limits from observations of qLMXB X7 in 47Tuc~\cite{Heinke}.\label{fig:mrobs1}}
\end{figure*}
\clearpage

\section{Neutron star observations}
\label{sec:observe}
\begin{figure*}
\plotone{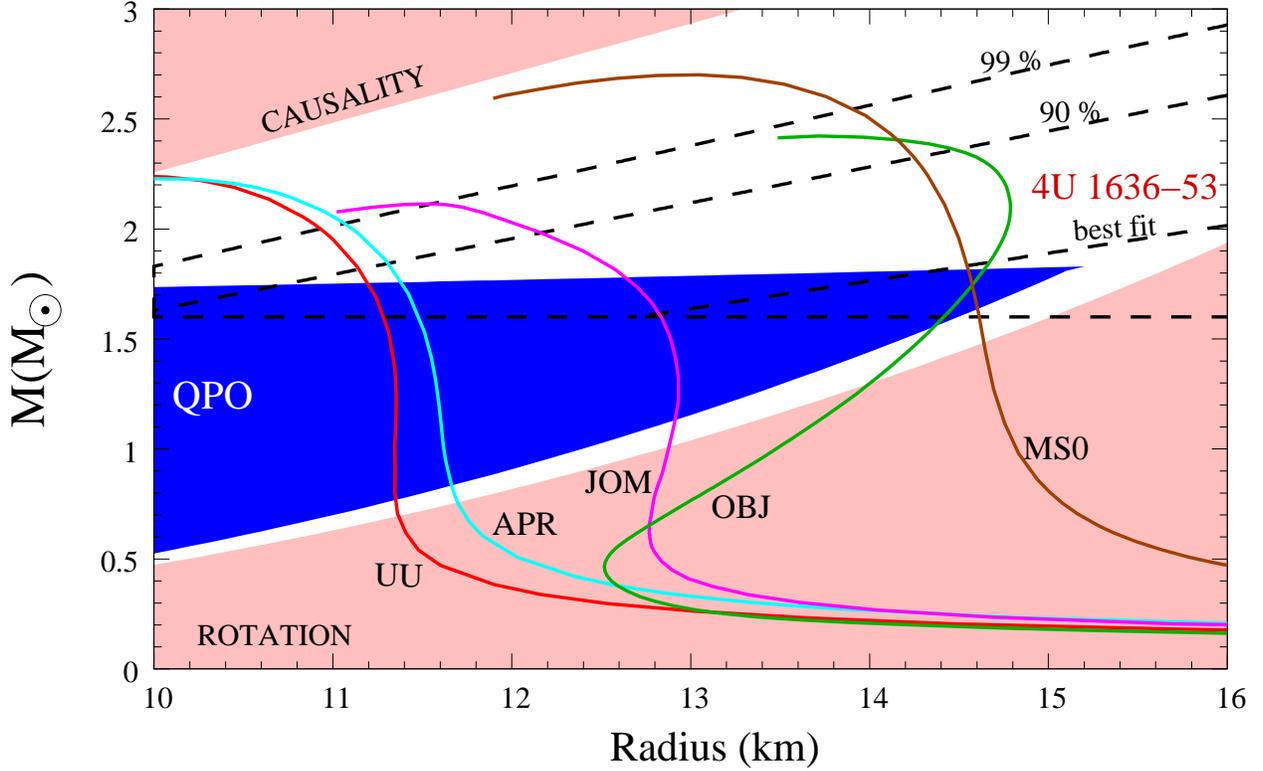}
\caption{Constraints for rapidly rotating stars: The blue wedge is the region allowed by kHz QPO oscillations in 4U 1728-34~\cite{Mill,barret} assuming a typical spin frequency of 300 Hz, while the dashed wedges are the regions allowed by compactness values $M/R$=0.126 (best-fit), 0.163 (90\%) and 0.183 (99\%) inferred from X-ray burst oscillations of LMXB 4U 1636-53~\cite{nss}. We assume its mass to be 1.6$M_{\odot}$. All mass-radius curves are for stars rotating with spin freqency $\nu$=300 Hz (period $P$=3.3ms). See text for details.\label{fig:mrobs2}}
\end{figure*}

In  Figs.\ref{fig:mrobs1},\ref{fig:mrobs2} we show several constraints on a neutron star's mass and radius that follow from general theoretical principles and observations of neutron star phenomena.~Fig.~\ref{fig:mrobs1} is relevant for static or slowly rotating stars.~A measurement of the blackbody radiation radius $R_{\infty}$=16.5km~\cite{trump04} and distance estimate $d$=117pc for the thermally emitting neutron star RXJ 1856.5-3754
~\cite{WL02} yields the red allowed region, which is a lower bound on the radius since the blackbody radiator has the smallest emitting area for a given  luminosity.~Observations of glitches in the spin-down of the Vela pulsar place a one-parameter constraint on the fractional moment of inertia in the crust $\Delta I/I\approx 0.014$~\cite{Link}. This parameter is the pressure at the core-crust interface, which takes typical values in the range $0.25<P/({\rm MeV/fm^3})<0.65$. The upper and lower limits translate to excluded regions in the upper left region of Fig.\ref{fig:mrobs1} marked Vela-glitch 1 and 2 respectively. Note that the region marked Vela-glitch 2, if taken at face value, rules out even a 1.6$M_{\odot}$ star for the APR and UU EoS. The region marked Vela-glitch 1 still allows the APR and UU EoS up to 1.8$M_{\odot}$.~Causality of the EoS excludes  a (partially overlapping) smaller pink region in the upper left corner.~A scaling relation between mass and radius for the minimum allowed spin period ~\cite{lps}, combined with the highest observed spin-frequency (716Hz)  of J1748-2446ad~\cite{Hess}, excludes the region marked "rotation". 90\% confidence limits on the radiation radius for the thermal source X7 in the globular cluster 47Tuc~\cite{Heinke}, arising from atmospheric modelling that includes surface gravity effects consistently, are shown by the dashed yellow lines. \"Ozel's estimate~\cite{ozel} of mass and radius based on a red-shift measurement $z$=0.35 ~\cite{Cottam} and an assumed atmosphere of accreted hydrogen for EXO 0748-676 gives the solid black line, with 1$\sigma$ and 2$\sigma$ limits displayed.

We have plotted the mass-radius curve from the JOM and OBJ EsoS as well as the APR, UU and MS0 EsoS. Among these, only MS0 and OBJ lie well inside the red allowed region for a star$\simeq 1.4M_{\odot}$. A smaller distance, still within the above $\pm 12$ uncertainty limits, would also allow the JOM EoS, provided the mass of RXJ 1856.5-3754 is $\simeq 1.8M_{\odot}$. The relatively soft equations of state such as APR and UU do not satisfy this constraint coming from RXJ 1856.5-3754 unless the mass of this object $\geq 2.0M_{\odot}$ and the glitch constraint is ignored. Although MS0 and OBJ appear promising in this light, as argued before, they are theoretically incomplete without higher-order terms, from the standpoint of an effective field theory. The addition of extra quartic terms to the Lagrangian in the OBJ model, which is essential for the self-consistency of the approximations and the truncation scheme, softens the EoS considerably. This yields the JOM EoS, which comes closest to satisfying all constraints. It is noteworthy that the APR, UU, OBJ and JOM mass-radius curves are also consistent with recently determined bounds (not shown in the figure) on the radiation radius of neutron stars in the globular cluster M13~\cite{Gendre1} and $\omega$-Centauri~\cite{Gendre2} while MS0 is not. Also, in addition to the JOM EoS, the APR and UU EoS are consistent at the 2$\sigma$ level, but not at the 1$\sigma$ level, with observational constraints set by X-ray bursts in EXO 0748-646 .

\vskip 0.2cm

Therefore, accepting all these observations as accurate, the OBJ and JOM EsoS are unique among the stiff EsoS considered here, in that they are most likely to satisfy all constraints from the afore-mentioned observations. Unlike the OBJ EoS however, the JOM EoS can also satisfy constraints from laboratory experiments, as demonstrated in $\S$\ref{sec:model}. The JOM EoS also implies that the mass of the neutron star RXJ 1856.5-3754 has to be $\simeq 1.8M_{\odot}$ and that of the qLMXB X7 has to be $\geq 1.6M_{\odot}$.

\clearpage

In Fig.\ref{fig:mrobs2}, we display mass-radius curves for rapidly rotating stars with the same EsoS as in the previous figure.~Constraints from kHz quasi-periodic oscillations (QPOs) in the bursting neutron star 4U 1728-34~\cite{Mill,barret} and practical upper limits on spin-frequencies of neutron stars with hadronic EsoS~\cite{lps} yield the wedge-shaped allowed region in blue. We have assumed a fairly typical QPO spin-frequency of 300 Hz in obtaining this constraint. We also employ compactness constraints obtained from the analysis of Nath et al. (2002) which assume a two-spot model to explain the X-ray burst oscillations of LMXB 4U 1636-53. Their best-fit value for compactness is $M/R$=0.126 assuming a spin-frequency of 290 Hz. With an assumed mass of 1.6$M_{\odot}$~\cite{casa} for this object, we obtain the smallest of the allowed triangular region bounded by dashes in the central right region of the figure. With 90\% ($M/R$=0.163) and 99\% ($M/R$=0.183) confidence-level constraints, we obtain the larger triangular regions (also demarcated in Fig.\ref{fig:mrobs2} by dashed lines). We plot the mass-radius curves for rotating configurations with $\nu$=$300$Hz in order to make a fair comparison to the constraints. We observe that the rotating APR and UU configurations cannot satisfy the combination of constraints if the best-fit compactness value is used but the MS0, OBJ and JOM EsoS can. However, if we choose a lower mass $M$=1.44$M_{\odot}$ or higher confidence limits for the compactness of 4U 1636-53, the APR and UU EoS can also satisfy this constraint. Accretion argues against having a mass smaller than 1.6$M_{\odot}$~\cite{Giles}. Thus, an accurate mass measurement of this object, although complicated due to its accreting nature, would be very desirable.

We have not attempted here to exhaustively survey all the EsoS that have been investigated in the literature and compared to observations, nor utilize all possible observational constraints. There are a variety of EsoS based on relativistic extensions of potential models~\cite{stein}, simple parameterized EsoS~\cite{pal1} and relativistic field theories with density dependent couplings~\cite{kbt}, only a few of which (eg. PAL1, MPA1 mentioned in the introduction) can achieve consistency with a majority of observational and experimental data. There are constraints from neutron star seismology~\cite{Duncan} which are satisfied only with the softest EoS possible in our model, one that cannot generate a
2.0$M_{\odot}$ maximum mass for a neutron star. While improved limits from future observations would naturally prove more definitive, the above discussion is meant to demonstrate that, at present, only an equation of state that is quite stiff at high density and moderately so near saturation density can satisfy a preponderance of astrophysical and laboratory constraints. The JOM EoS, based on a relativisitic mean field theory with scale-breaking and symmetry breaking inspired from QCD, provides one such promising example.

\section{Conclusions}
\label{sec:conc}

We have examined a mean field model for dense nuclear and neutron-rich matter, with direct application to neutron star interiors. The model is derived from the Skyrme Lagrangian, with the inclusion of the dilaton VEV performing a dual role: (i) to mock up scale-breaking in QCD and (ii) to dial the interaction between the density-dependent vector-meson mean fields and the skyrmion fluid which makes up the dense medium, thus achieving self-consistency. We included quartic self-interactions for the vector mesons to correct an unnaturally rapid rise of the compressibility and symmetry energy at densities just above saturation. The resulting equation of state is in better agreement with laboratory constraints on the compressibility and symmetry energy. At densities much above saturation, the equation of state is stiff; consequently it yields a large maximum mass for a neutron star. This is particularly interesting given accumulating data that points to the existence of neutron stars with $M\sim 2M_{\odot}$, though corresponding systematic errors tend to be larger than is the case for stars with $M\sim 1.4M_{\odot}$.

The pressure and symmetry energy of dense matter, as well as the mass-radius curves for neutron stars depend sensitively on the (natural) values of the quartic couplings. Increasing the $\omega$-meson self-coupling reduces the compressibility as well as the symmetry energy near saturation density considerably, while increasing the $\rho$-meson self-coupling stiffens the EoS of neutron-rich matter at high density, thereby increasing the maximum mass. The maximum neutron star mass in our model lies between 1.8 to 2.0$M_{\odot}$ for static configurations with corresponding radius at maximum mass between 10.5-11.5km. For rapidly rotating stars, these values increase to 2.0-2.3$M_{\odot}$ and 11.5-12.5 km respectively. The minimum spin period for the maximum mass star lies between $0.66~{\rm ms}$ and $ 0.71~{\rm ms}$. These macroscopic effects are traceable to changes in the dilaton VEV, which effectively controls the Skyrmion size through its interactions with the density-dependent vector-meson mean fields, thereby determining the stiffness of the EoS, and highlighting the role of scale-breaking in our model.

Confronting our relatively stiff EoS with a set of observational constraints on neutron star mass and radius, we find encouraging agreement. Including commonly used EsoS based on extrapolations of microscopic models of the nucleon-nucleon interaction, we infer that matter is likely to be stiffer at supranuclear densities than such models would suggest. Therefore, an accurate and simultaneous determination of mass and radius for the more 'extreme' neutron stars, viz., those presently suggestive of high mass and large radius, will be particularly valuable. In addition, constraints from collective flow data in heavy-ion collisions and the isospin dependence of the strong interaction near saturation density are proving to be valuable benchmarks for dense matter studies.

%\section{Acknowledgments}
%\label{sec:ack}
%\vskip -0.4cm

\acknowledgments
This research was supported by the Department of Atomic Energy (DAE) of the Government of India and US-DOE grant DE-FG02-93ER40756. R. O. is supported by the Natural Science and Engineering Research Council of Canada (NSERC). We are grateful to K. Mori, J. M. Lattimer and M. Prakash for discussions; A. W. Steiner for providing some of the equations of state used in this work; and N. Stergioulas for making the $RNS$ code easily available to the community.

\clearpage
\begin{deluxetable}{rrrrrr} 
\label{fittable}
\tablecolumns{8} 
\tablewidth{0pc} 
\tablecaption{Fit parameters for the JOM model for various $\xi\neq$ 0. The OBJ model has $\xi$=0. In both cases, $K$=240MeV, B.E.=$-$16MeV, ${\rm E}_{sym}$=32MeV, $g_{\rho}$=4.917}
\tablehead{ \colhead{$\xi$} & \colhead{$g_{\omega}$}   & \colhead{$a_1$}    & \colhead{$a_2$} & \colhead{$a_3$}    & \colhead{$a_4$}}
\startdata 
0.0  & 7.54 & -1.699 & -25.780 & 29.237 & -11.464\\
0.1  & 6.67   & -1.960 & -27.051 & 29.798 & -11.502\\
0.2  & 6.05 & -2.260 & -28.700 & 30.674 & -11.636\\
0.3  & 5.57 & -2.602 & -30.747 & 31.878 & -11.870\\ 
\enddata 
\end{deluxetable}


\begin{thebibliography}{}


\bibitem[Akmal et~al 1998]{apr} Akmal, A., Pandharipande, V. R., \& Ravenhall, D. G. 1998, Phys. Rev. C, 58, 1804.

\bibitem[Alford et al. 2007]{Alf} Alford, M. G., Blaschke, D., Drago, A., Klahn, T., Pagliara, G., \& Schaffner-Bielich, J. 2007, Nature 445, E7.

\bibitem[Arnett \& Bowers 1977]{ab} Arnett, W. D. \& Bowers, R. L. 1977, ApJS, 33, 415.

\bibitem[ATNF Pulsar Catalogue]{ATNF} Australia Telescope National Facility Pulsar Catalogue (http://www.atnf.csiro.au/research/pulsar/psrcat/); see also Manchester, R. N., Hobbs, G. B., Teoh, A., \& Hobbs, M. 2005, ApJ, 129, 1993.

\bibitem[Barret et al. 2006]{barret} Barret, D., Olive, J.-F., \& Miller, M. C. 2006, MNRAS, 370, 1140.

\bibitem[Baym et al. 1971a]{BBP} Baym, G., Bethe, H. A. \& Pethick, C. J. 1971, Nucl. Phys. A, 175 , 225. 

\bibitem[Baym et al 1971b]{BPS} Baym, G, Pethick, C. J., \& Sutherland, P. 1971, 170, 299.

\bibitem[Casares et al. 2006]{casa} Casares, J., Cornelisse, R., Steeghs, D., Charles, P. A., Hynes, R. I., O'Brien, K., \& Strohmayer, T. E. 2006, MNRAS, 373, 1235.

\bibitem[Cottam et al. 2002]{Cottam} Cottam, J., Paerels, F., \& Mendez, M. 2002, Nature, 420, 51.
.
\bibitem[Danielewicz et al. 2002]{dll} Danielewicz, P., Lacey, R., \& Lynch, G. W. 2002, Science, 298, 1592.

\bibitem[Duncan 1998]{Duncan} Duncan, R. C. 1998, Astrophys. J. Lett. 498, L45.

\bibitem[Forest et al. 1995]{Forest} Forest, J. L., Padhanripande, V. R., \& Friar, J. L. 1995, Phys. Rev. C, 52, 568.

\bibitem[Furnstahl et al. 1996]{Furn} Furnstahl, R. J., Serot, B. D., \& Tang, H.-B. 1996, Nucl. Phys. A, 598, 539.

\bibitem[Gendre et al. 2003a]{Gendre1} Gendre, B., Barret, B., \& Webb, N. A. 2003, A \& A, 401, L33.

\bibitem[Gendre et al. 2003b]{Gendre2} Gendre, B., Barret, B., \& Webb, N. A. 2003, A \& A, 400, 521.

\bibitem[Giles et al. 2002]{Giles} Giles, A. B., Hill, K. M., Strohmayer, T. E., \& Cummings, N. 2002, Astrophys. J., 568, 279.

\bibitem[Heinke et al. 2006]{Heinke}Heinke, C. O., Rybicki, G. B., Narayan, R., \& Grindlay, J. E. 2006, ApJ, 644, 1103.

\bibitem[Hessels et al. 2006]{Hess} Hessels, J. W. T. 2006, Science, 311, 1801.

\bibitem[Hewish et al. 1968]{Bell} Hewish, A., Bell, S. J., Pilkington,
J.~D.~H., Scott, P.~F., \& Collins, R.~A. 1968, Nature, 217, 709.

\bibitem[Ho et al. 2007]{Ho} Ho, W. C. G., Kaplan, D. L., Chang, P.; van Adelsberg, M.; Potekhin, A. Y. 2007, MNRAS, 375, 821.

\bibitem[Jaikumar \& Ouyed 2006]{jo} Jaikumar, P. \& Ouyed, R. 2006, ApJ, 639, 354.

\bibitem[K\"albermann 1997]{kal} K\"albermann, G. 1997, Nucl. Phys. A, 612, 359.

\bibitem[Kl\"ahn et~al. 2006]{kbt} Kl\"ahn, T., Blaschke, D., Typel, S., $et~al.$ 2006, Phys. Rev. C, 74, Issue 3, 035802.

\bibitem[Lattimer \& Prakash 2001]{lp} Lattimer, J. M. \& Prakash, M. 2001, ApJ, 550, 426.

\bibitem[Lattimer \& Prakash 2004]{lps} Lattimer, J. M. \& Prakash, M. 2004, Science 304, 536.

\bibitem[Lattimer \& Prakash 2005]{lp05} Lattimer, J. M. \& Prakash, M. 2005,
Phys. Rev. Lett., 94, 1101.

\bibitem[Lattimer \& Prakash 2007]{lp2} Lattimer, J. M. \& Prakash, M. 2007, Phys. Rept.442, 109.

\bibitem[Li \& Steiner 2006]{ls2006} Li, B.-A. \& Steiner, A. W. 2006, Phys. Lett. B, 642, 436

\bibitem[Link et al. 1999]{Link} Link, B., Epstein, R. I., \& Lattimer, J. M. 1999, Phys. Rev. Lett. 83, 3363.

\bibitem[Miller et al. 1998]{Mill} Miller, M. C., Lamb, F. K., \& Saltis, D. 1998, 508, 791.

\bibitem[M\"uther et al. 1987]{mpa1} M\"uther, H., Prakash, M., \& Ainsworth, T. L. 1987, Phys. Lett. B, 199, 469.

\bibitem[M\"uller \& Serot 1996]{ms} M\"uller, H. \& Serot, B. D. 1996, Nucl. Phys. A, 606, 508.

\bibitem[Nath et al. 2002]{nss} Nath, N. R., Strohmayer, T. E., \& Swank, J. H. 2002, ApJ, 564, 353.

\bibitem[Nice et al. 2007]{Nice} Nice, D. J. et al., at {\it Montreal 2007: 40 years of Pulsars}.

\bibitem[Ouyed \& Butler 1999]{ob} Ouyed, R. \& Butler, M. 1999, ApJ, 522, 453.

\bibitem[\"Ozel 2006]{ozel} \"Ozel, F. 2006, Nature, 441, 1115.

\bibitem[Page \& Reddy 2006]{page} Page, D. \& Reddy, S. 2006, Ann. Rev. Nucl. Part. Sci. 56, 327.

\bibitem[Pearson et al. 2006]{Pearson} Pearson, K. J., Hynes, R. I., Steeghs, D., Jonker, P. G., Haswell, C. A., King, R. A., O'Brien, K., Nelemans, G., \& Mendez, M. 2006, ApJ, 648, 1169.

\bibitem[Prakash et al. 1988a]{pal1}  Prakash, M., Ainsworth, T. L., \& Lattimer, J. M. 1988a, Phys. Rept. 442, 109.

\bibitem[Prakash et al. 1988b]{pal2}  Prakash, M., Ainsworth, T. L., \& Lattimer, J. M. 1988b, Phys. Rev. Lett. 61, 2518.

\bibitem[Pudliner et al. 1995]{Pud} Pudliner, B. S., Pandaharipande, V. R., Carlson, J., \& Wiringa, R. B. 1995, Phys. Rev. Lett., 74, 4396.

\bibitem[Rami et al. 2000]{Rami} Rami, F. et al. 2000, Phys. Rev. Lett, 84, 1120.

\bibitem[Ransom et al. 2005]{ransom05} Ransom, S. R., Hessels, J. W. T., Stairs, I. H., Freire, P. C. C, Camilo, F., Kaspi, V. M. \& Kaplan, D. 2005, Science, Vol. 307, Issue 5711, 892.

\bibitem[Schecter \& Weigel 2000]{Schecter} Schecter, J. \& Weigel, H. 2000, review article published in INSA-Book-2000; arXiv:hep-ph/9907554.

\bibitem[Sedrakian 2006]{Sed} Sedrakian, A. 2006, Prog. Part. and Nucl. Phys.,
 Vol. 58, Issue 1, 168.

\bibitem[Skyrme 1961]{sky} Skyrme, T. H. R, 1961, Proc. Roy. Soc. Lond. A, 262, 237.

\bibitem[Starodubsky \& Hintz 1994]{star} Starodubsky, V. E. \& Hintz, N. M. 1994, Phys. Rev. C, 49, 2118.

\bibitem[Steiner 2006]{AS} Steiner, A. W. 2006 , Phys. Rev. C, 74, 045808.

\bibitem[Steiner et al. 2005]{stein} Steiner, A. W., Prakash, M, Lattimer, J. M., \& Ellis, P. J. 2005, Phys. Rept., 411, 325.

\bibitem[Stergioulas \& Freidman 1995]{Ster95} Stergioulas, N. \& Friedman, J. L. 1995, ApJ, 444, 306

\bibitem[Stergioulas \& Freidman 1998]{Ster98} Stergioulas, N. \& Friedman, J. L. 1998, ApJ, 492, 301.

\bibitem[Tr\"umper et al. 2004]{trump04}Trumper, J. E., Burwitz, V., Haberl, F. \& Zavlin, V. E. (2004), Nucl. Phys. Proc. Suppl., 132, 560.

\bibitem[Tsang et al. 2004]{tsang} Tsang, M. B. et al. 2004, Phys. Rev. Lett. 92, 062701.

\bibitem[Walter \& Lattimer 2002]{WL02} Walter, F. M. \& Lattimer, J. M. 2002, ApJ, 576, 145.

\bibitem[Wiringa et al. 1988]{Wiringa} Wiringa,  R. B., Fiks, V., \& Fabrocini, A., 1988, Phys. Rev. C, 38, 1010.

\bibitem[Wiringa et al. 1995]{Stoks} Wiringa, R. B., Stoks, V. G. J., \& Schiavilla, R. 1995, Phys. Rev. C, 51, 38.
\end{thebibliography}
\end{document}